\documentclass[aps,twocolumn,nofootinbib,preprintnumbers,floatfix,showpacs,prx]{revtex4-1}

\usepackage{amsmath,amssymb}
\usepackage{dcolumn}
\usepackage{braket}
\usepackage{dsfont}
\usepackage{feynmp}
\usepackage{hhline}
\usepackage{graphicx}
\usepackage{mathtools}
\usepackage{multirow}
\usepackage{xfrac}
\usepackage{paralist}
\usepackage{enumitem}
\usepackage{simplewick}
\usepackage[normalem]{ulem}
\usepackage[usenames,dvipsnames]{color}
\usepackage[colorlinks=true,citecolor=blue,linkcolor=magenta]{hyperref}

\newcommand{\rb}{\boldsymbol{r}}
\newcommand{\sB}{\boldsymbol{s}}

\DeclareMathOperator{\Tr}{\mathrm{Tr}}
\DeclareMathOperator{\mS}{\mathcal{S}}
\DeclareMathOperator{\mN}{\mathcal{N}}
\DeclareMathOperator{\mU}{\mathcal{U}}
\DeclareMathOperator{\mM}{\mathcal{M}}
\DeclareMathOperator{\mJ}{\mathcal{J}}
\begin{document}

\title{Finite-Temperature Scrambling of a Random Hamiltonian}
\author{Sagar Vijay}
\affiliation{Department of Physics, Massachusetts Institute of Technology, Cambridge, MA 02139, USA}
\author{Ashvin Vishwanath}
\affiliation{Department of Physics, Harvard University, Cambridge, MA 02138, USA}
\begin{abstract}
We study the finite-temperature scrambling behavior of a quantum system described by a Hamiltonian chosen from a random matrix ensemble.  This effectively (0+1)-dimensional model admits an exact calculation of  various ensemble-averaged out-of-time-ordered correlation functions in the large-$N$ limit, where $N$ is the Hilbert space dimension.  { For a Hamiltonian drawn from the Gaussian unitary ensemble, we calculate the ensemble averaged OTOC at all temperatures}. In addition to an early time quadratic growth of the averaged out-of-time-ordered commutator (OTOC), we determine that the OTOC saturates to its asymptotic value as a power-law in time, with an exponent that depends both on temperature, and on one of four classes of operators appearing in the correlation function, that naturally emerge from this calculation.  
Out-of-time-ordered correlation functions of operators that are distributed around the thermal circle take a time $t_{s}\sim \beta$ to decay at low temperatures.  We generalize these exact results, by demonstrating that out-of-time-ordered correlation functions averaged over any ensemble of Hamiltonians that are invariant under unitary conjugation $H \rightarrow \mU H \mU^{\dagger}$, exhibit power-law decay to an asymptotic value. We argue that this late-time behavior is a generic feature of unitary dynamics with energy conservation.  
Finally, by comparing the OTOC with a commutator-anticommutator correlation function, we examine whether there is a time window over which a typical Hamiltonian behaves as a ``coherent scrambler" in the language of Ref. \cite{Kitaev_IAS, Kitaev_Suh}. 
\end{abstract}
\maketitle

The propagation of quantum information, and the thermalization of local observables in a quantum many-body system undergoing unitary evolution have been subjects of much recent interest.   
Certain universal dynamical properties of quantum information spreading have been understood, such as the presence of ``speed limits" for information propagation \cite{lieb_robinson}, and scaling forms for the growth of the von Neumann entanglement entropy \cite{CardyCalabrese2005EntanglementEvolution,
CalabreseCardyQuenchReview,LiuSuh2014,Asplund,CasiniLiuMezei2015,NRVH2016, mezei2016entanglement,
SerbynPapicAbanin2013,HuseNandkishoreOganesyan2014}, which govern the approach to thermalization. 

Unitary evolution also gives rise to an effectively irreversible process of information scrambling, in which initially locally accessible quantum information becomes practically irretrievable.  Scrambling may be quantified by out-of-time-ordered correlation functions, such as the out-of-time-ordered commutator (OTOC)
\begin{align}
&C_{\beta}(t) \equiv \frac{1}{2}\Big\langle \big| \left[V(t), W(0)\right]\big|^{2}\Big\rangle_{\beta}\label{eq:C}
\end{align}
which measures the ``spreading" of the operator $V(t) = \mU(t)V(0)\mU(t)^{\dagger}$ under Heisenberg evolution.  While {out-of-time-ordered correlations} were introduced to study quasi-classical approximations to superconductivity~\cite{larkin}, they have recently been used as a diagnostic for quantum chaos \cite{kitaev_fundamental_physics_symposium, shenker_stanford_1, shenker_stanford_2, maldacena2016, Aleiner, 
Stanford_Chaos, Sachdev_CriticalFermiSurface, Altman, gu2016local, RobertsSwingle2016, chowdhury2017onset, Sachdev_DiffusiveMetals, roberts2015diagnosing, bohrdt2016scrambling,luitz2017information,leviatan2017quantum,dora2016out}.  The exact form of certain out-of-time-ordered correlations has also been studied in random quantum circuits \cite{Op_Spreading_1, Op_Spreading_2, Op_Spreading_3, Op_Spreading_4, Floquet_Chalker}, and these correlations have been used to characterize slow dynamics in the presence of disorder and in the many-body localized 
phase~\cite{swingle2017slow, ForthcomingGriffiths, huang2016out, Chen2016, chen2016universal, 
he2017characterizing, fan2017out, chapman2017classical}. 
Experiments studying out-of-time-ordered correlations \cite{Tarun, Swingle, Yao} have been conducted~\cite{AMRey,
li2016measuring, wei2016exploring}. 

Significant progress in understanding information scrambling has been made by investigating out-of-time-ordered correlations in (0+1)-dimensional systems, such as the Sachdev-Ye-Kitaev (SYK) model \cite{SYK_1, SYK_2}, a model of $2\mN$ real (Majorana) fermions with all-to-all, random four-fermion interactions.  
The SYK model develops an approximate conformal symmetry at low energies \cite{maldacena_stanford}, and exhibits chaotic behavior, with the OTOC growing exponentially $C_{\beta}(t) \sim \mN^{-1}\exp\left[\lambda_{L}t\right]$, with a Lyapunov exponent $\lambda_{L} = 2\pi /\beta$ that saturates a known bound \cite{maldacena2016}.  These observations have motivated the conjecture that the SYK model admits a dual gravitational description.

Motivated by studies of scrambling in the SYK model, we study the behavior of out-of-time-ordered correlation functions at finite temperature, in a system described by a Hamiltonian chosen from a random matrix ensemble.  Such a Hamiltonian will be highly non-local, and it is instructive to compare scrambling in this effectively (0+1)-dimensional system, with the SYK model and with other chaotic quantum systems.  We present exact calculations of ensemble-averaged out-of-time-ordered correlation functions, and argue that certain features of the OTOC within a late-time window are universal for unitary dynamics with energy conservation.   {We note that unlike the SYK model, which allows for all-to-all $q$-fermion couplings (typically $q=4$), with $q$ held fixed as the number of degrees of freedom becomes large, a typical random Hamiltonian will admit arbitrarily long-ranged interactions. }

We first consider Hamiltonians drawn from the Gaussian unitary ensemble (GUE), and present an exact calculation of the ensemble-averaged OTOC at any temperature, in the large-$N$ limit, where $N$ is the dimension of the Hilbert space of the system.  From our calculation, we identify a regime between the relaxation time of typical observables, and a scrambling time, where the OTOC grows quadratically $\overline{C_{\beta}(t)}\sim J^{2}t^{2}$, at any temperature.  The OTOC saturates to its asymptotic value as a power-law, with an exponent that  both depends on temperature and on one of four classes of operators appearing in the correlation function, that naturally emerge from this exact calculation.  While the OTOC reaches an $O(1)$ value at a ``scrambling time" $t_{s}\sim 1/J$, out-of-time-ordered correlation functions that are distributed on the thermal circle 
exhibit power-law decay over a time $t_{s}\sim \beta$ at low temperatures.  
{ We note that a previous calculation of the out-of-time-ordered correlation function for a GUE Hamiltonian -- but restricted to \emph{infinite temperature}, and to a particular choice of operators $V$, $W$ -- was shown in Ref. \cite{Yoshida}.  In our calculation, we obtain the analytic expressions for the full time dependence of various out-of-time-ordered correlation functions at all temperatures in a large-$N$ limit, which reveals several interesting features -- such as the late-time dependence of the OTOC on the choice of operators -- that we discuss at length.  } 

We generalize our exact results by demonstrating that the OTOC, when averaged over any ensemble of Hamiltonians that is invariant under unitary conjugation $H\rightarrow \mU H \mU^{\dagger}$, will (i) saturate as a power law with an exponent that is bounded from below by a function of the density of states and (ii) scramble over a time $t_{s}\sim O(1/J)$, where $J$ is of the order of the width of the eigenvalue spectrum of the Hamiltonian. Therefore, a typical Hamiltonian from this ensemble will scramble much faster than a chaotic quantum system.  We note that the  $U(N)$ symmetry in this  ensemble of Hamiltonians -- with $N$, the dimension of the Hilbert space -- contrasts with the small symmetry group for a typical quantum many-body system. 
Nevertheless, we argue that the power-law saturation of out-of-time-ordered correlation functions at sufficiently late times -- longer than the scrambling time, but shorter than the inverse level spacing -- is a generic feature of unitary dynamics with energy conservation. %


We conclude by defining a criterion for ``coherent scrambling", as originally introduced in Ref. \cite{Kitaev_IAS, Kitaev_Suh}.  A quantum many-body system exhibits a coherent scrambling regime, if certain out-of-time-ordered correlation functions can be regarded as matrix elements of a unitary scattering matrix.  This perspective is motivated by holographic calculations, in which certain out-of-time-ordered correlation functions may be regarded as time-ordered correlations in a ``bulk" description of the  system of interest.  We define a scattering matrix with this property and show that unitarity requires that the commutator/anti-commutator correlation function 
\begin{align}\label{eq:D_intro}
D_{\beta}(t) = \frac{1}{2}\Big\langle [V(t), W(0)]\{V(t), W(0)\}\Big\rangle_{\beta}
\end{align}
satisfies $|D_{\beta}(t)| \gg C_{\beta}(t)$, for times that are longer than the relaxation time of the system ($t_{r}$), when time-ordered correlation functions have decayed to their asymptotic values.  For a random GUE Hamiltonian, this condition is satisfied by the ensemble-averaged correlation functions at low temperatures, and over times $t\lesssim t_{s}$, but this condition is satisfied over a much longer interval by the SYK model at low temperatures.

We also derive a stronger condition for coherent scrambling, that unitarity of the scattering matrix requires that there is {some} choice of operators $V$ and $W$, for which  
\begin{align}\label{eq:intro_unitarity}
C_{\beta}(t) \sim \frac{|D_{\beta}(t)|^{2}}{\langle W(0) V(t)^{2}W(0)\rangle_{\beta}} \hspace{.2in} (t \gtrsim t_{r}).
\end{align}
We emphasize that this condition may not necessarily be satisfied for an arbitrary choice of operators $V$, $W$.  We define the \emph{coherent scrambling time} $t_{c}$ as the timescale when this condition is satisfied, and show that for a GUE Hamiltonian, $t_{c}$ is of the same order as the scrambling time of the system, at low temperatures.  The condition (\ref{eq:intro_unitarity}) seems to be violated by simple out-of-time-ordered correlation functions of Majorana fermions in SYK model. 

{ In an independent forthcoming work, the entanglement features of the time-evolution operator for a random Hamiltonian have been evaluated using similar techniques \cite{Yi-Zhuang_Yingfei}. \\}

\section{Finite Temperature Scrambling of a GUE Hamiltonian}
We begin by presenting the exact results of our finite temperature calculation of the out-of-time-ordered commutator for a Hamiltonian draw from the Gaussian unitary ensemble.  Let $H$ be an $N\times N$ Hermitian Hamiltonian drawn from this ensemble, which is completely specified by the following moments
\begin{align}
\overline{H_{ab}} = 0,\hspace{.43in} \overline{H_{ab}H_{cd}} = \frac{J^{2}}{N}\delta_{ad}\delta_{bc}.
\end{align}
Here, the line $\overline{\cdots}$ denotes an ensemble average.  To diagnose scrambling, we compute the ensemble-averaged out-of-time-ordered commutator  $\overline{C_{\beta}(t)}$, as defined in (\ref{eq:C}), 
where $\langle\cdots\rangle_{\beta}$ indicates a thermal average with respect to the Hamiltonian, at temperature $T = \beta^{-1}$.  The time-evolved operator $V(t)$ is given by Heisenberg evolution $V(t) = e^{-iHt}V(0)e^{iHt}$. 

\begin{table*}
\begin{tabular}{ | c | c | c | c | c | c | c | }
  \cline{4-7} 
 \multicolumn{3}{c|}{} & \multicolumn{2}{c|}{$\begin{array}{c}\\\overline{C_{\beta}(t)} \sim\\\\\end{array}$} & \multicolumn{2}{c|}{$\begin{array}{c}\\\displaystyle\overline{C_{\beta}(\infty)} - \overline{C_{\beta}(t)} \sim\\\\\end{array}$} \\
 
 \hline
 \multicolumn{2}{|c|}{$\boldsymbol{V(0)}$ {\bf and} $\boldsymbol{W(0)}$} & {\bf Example} & $\begin{array}{c}\beta J \rightarrow\infty\\ Jt\rightarrow 0\end{array}$ & $\begin{array}{c}\beta J\rightarrow0\\Jt\rightarrow 0\end{array}$ & $\begin{array}{c}\beta J\rightarrow\infty\\Jt\rightarrow \infty\end{array}$ & $\begin{array}{c}\beta J\rightarrow0\\Jt\rightarrow \infty\end{array}$\\ 

 \hline\hline
 \multicolumn{2}{|c|}{Disjoint} & $\begin{array}{c} V(0) = X_{\rb}\\ W(0) = Z_{\sB} \end{array}$ & \multicolumn{2}{c|}{$J^{2}t^{2}$}  & $\displaystyle (Jt)^{-9/2}{\cos(2Jt)\cos^{3}\left(\displaystyle\frac{\pi}{4} + 2Jt\right)}$ & $ \displaystyle (Jt)^{-6}{\displaystyle\cos^{4}\left(\frac{\pi}{4} + 2Jt\right)}$ \\
 
 \hline
 \multirow{8}{3.5em}{Overlap} &  $\begin{array}{c}\\VW = 0\\\\\end{array}$ & $\begin{array}{c}V(0) = P^{(+)}_{\rb}Z_{\sB'}\\ W(0) = P^{(-)}_{\rb}Z_{\sB}\end{array}$ & \multicolumn{2}{c|}{$''$} & \multicolumn{2}{c|}{$ \displaystyle (Jt)^{-3}{\displaystyle\cos^{2}\left(\frac{\pi}{4} + 2Jt\right)}$} \\
 
 \cline{2-7}
 & $\begin{array}{c}\\ \Tr(V\,W)= 0,\\ \Tr(V^{2}W^{2})\ne 0\\\\\end{array}$ & $\begin{array}{c}V(0) = Z_{\rb_{1}}P^{(+)}_{\rb_{2}}\\ W(0) = P^{(+)}_{\rb_{2}}Z_{\rb_{3}}\end{array}$ & \multicolumn{2}{c|}{$''$}  & \multicolumn{2}{c|}{$''$} \\
 
  \cline{2-7}
& $\begin{array}{c}\\\Tr(V\,W)\ne 0\\\\\end{array}$ & $\begin{array}{c}V(0) = X_{\rb}\\ W(0) = X_{\rb}\end{array}$ & \multicolumn{2}{c|}{$''$}  & $ \displaystyle {(Jt)^{-5/2}}{\cos(2Jt)\cos\left(\displaystyle\frac{\pi}{4} + 2Jt\right)}$ & $\displaystyle{(Jt)^{-4}}{\cos^{2}\left(\displaystyle\frac{\pi}{4} + 2Jt\right)}$ \\
 \hline
 \end{tabular}
 \caption{{\bf Early and Late-Time Behavior of $\overline{C_{\beta}(t)}$}.   The out-of-time-ordered commutator generically grows quadratically at early times, and exhibits power-law saturation at late times $(1 \ll Jt \ll N)$, which depends on the choice of initially commuting operators $V$ and $W$. Examples of the four classes of operators are indicated, with $X_{\rb}$ and $Z_{\rb}$, denoting Pauli operators at site $\rb$ and $P^{(\pm)}_{\rb} \equiv (1\pm Z_{\rb})/2$. 
We have not indicated the temperature dependence of the $O(1)$ constants appearing in the early or late-time behavior of the OTOC.  }\label{table:C_Table}
\end{table*}

\subsection{Out-of-Time-Ordered Correlation Functions\\ in the Large-$N$ Limit}
For simplicity, we take $V$ and $W$ to be traceless Hermitian operators, that initially commute $[V(0), W(0)] = 0$.  In the Supplemental Material \cite{SM}, we calculate the out-of-time-ordered commutator, after averaging over the Gaussian ensemble for $H$.  We find that to leading order in the large-$N$ limit ($N\rightarrow\infty$), that 
\begin{widetext}
\begin{align}\label{eq:C_Complete}
\overline{C_{\beta}(t)} \,\,= \,\,&\frac{\Tr\left( V^{2}\right) \Tr\left( W^{2}\right)}{{N^{2}}}\left[1 -  \left(\frac{I_{1}(2iJt)}{iJt}\right)^{2}\right]
+ \frac{\Tr(V^{2}W^{2})}{N}\left(\frac{I_{1}(2iJt)}{iJt}\right)^{2}\left[ 1 - \mathrm{Re}\left\{\frac{\beta}{\beta + it}\frac{I_{1}(2J(\beta + it))}{I_{1}(2\beta J)}\right\}\frac{I_{1}(2iJt)}{iJt}\right]\nonumber\\
&- \frac{\Tr(V\,W)^{2}}{N^{2}}\frac{\beta }{I_{1}(2\beta J)} \frac{I_{1}(2iJt)}{it}\,\mathrm{Re}\left\{\frac{I_{1}(2iJt)}{iJt} F(\beta J + iJt, iJt) + \frac{I_{1}(2\beta J + iJt)}{ \beta J + iJt}F(iJt, iJt)\right\} + O(N^{-1})
\end{align}
\end{widetext}
where $I_{1}(z)$ is the modified Bessel function of the first kind, and the traces are taken over all states in the $N$-dimensional Hilbert space.  Here, the function
\begin{align}
F(z,w) \equiv \int dx\, dy \frac{(e^{-wx} - e^{-wy})(e^{-zx} - e^{-zy})}{(x-y)^{2}}\rho(x)\rho(y)\nonumber
\end{align}
where $\rho(x)$ is the Wigner semicircle distribution $\rho(x) \equiv (2\pi)^{-1}\sqrt{4-x^{2}}$ with $(|x| \le 2)$, which describes the probability distribution of a single eigenvalue of a Hamiltonian drawn from the Gaussian unitary ensemble with unit variance.  We note that fluctuations of $C_{\beta}(t)$ about its mean value are strictly zero in the large-$N$ limit, i.e. 
$\overline{C_{\beta}(t)^{2}} = \overline{C_{\beta}(t)}^{2} + O(N^{-1})$.  Therefore, our results are valid for a typical, random Hamiltonian in the thermodynamic limit.

The above expression reveals that the ensemble-averaged out-of-time-ordered commutator behaves distinctly, depending on whether 
\begin{enumerate}
\item $V(0)$ and $W(0)$ are \emph{disjoint} (non-overlapping) .
\item $V(0)$ and $W(0)$ have overlapping support.
\end{enumerate}
Within the second class of overlapping operators, the behavior of out-of-time-ordered quantities is further distinguished into three types, depending on whether the various traces of $V$ and $W$ appearing in (\ref{eq:C_Complete}) vanish.  A summary of the leading early- and late-time behavior of the out-of-time-ordered commutator, at both low and high temperatures, and for the different classes of operators are presented in Table \ref{table:C_Table}.  Representative examples of operators in the different classes are also indicated.

As shown in Table \ref{table:C_Table}, the out-of-time-ordered commutator grows quadratically in time, at early times and at any temperature, and for any choice of initially commuting operators, traceless operators $V(0)$ and $W(0)$. 
At long times the out-of-time-ordered commutator saturates to a value $\lim_{t\rightarrow\infty}\overline{C_{\beta}(t)} = N^{-2}\Tr(V^{2})\Tr(W^{2})$ as a power-law, as indicated in Table \ref{table:C_Table}, which depends sensitively on the choices of operators, and on the temperature.  We note that since we are working in the large-$N$ limit, the indicated power-law decay to saturation in Table \ref{table:C_Table} is valid for long times that are shorter than the inverse level spacing, i.e. $1 \ll Jt\ll N$.  

From (\ref{eq:C_Complete}), we may extract a scrambling time, by observing that $\overline{C_{\beta}(t)}$ exactly reaches its asymptotic value when $I_{1}(2iJt) = 0$. This condition is first satisfied at a time $t_{0} \approx 1.916/J$, and this provides an upper bound on the scrambling time at any temperature.  We note that the scrambling time, as extracted from the saturation of the OTOC, is very different from the scrambling time extracted from the {decay} of the correlation function 
\begin{align}\label{eq:thermal_circle}
F_{\beta}(t) = \Tr\left[\rho^{1/4}V(t)\rho^{1/4}W(0)\rho^{1/4}V(t)\rho^{1/4}W(0)\right]
\end{align}
where $\rho \equiv e^{-\beta H}/\Tr(e^{-\beta H})$ is the thermal density matrix.  Operators in this  correlation function are evenly distributed on the thermal circle. Analyticity of this correlation function has been used to derive bounds on the Lyapunov exponent for chaotic quantum systems \cite{maldacena2016}.

For $V$ and $W$, chosen to be initially disjoint, traceless operators, we observe that the ensemble-averaged correlation function (\ref{eq:thermal_circle}) is given by
\begin{align}
\overline{\frac{F_{\beta}(t)}{F_{\beta}(0)}} = \left| \frac{\beta\cdot I_{1}(2iJt - ({\beta J}/{2}))}{(4it - \beta)\cdot I_{1}(\beta J/2)}\right|^{4} + O(N^{-1})
\end{align}
so that at low temperatures and sufficiently long times
\begin{align}\label{eq:t6}
\overline{\frac{F_{\beta}(t)}{F_{\beta}(0)}} =\left(\frac{\beta }{4 t}\right)^{6}  + O(\beta/t)^{7}\hspace{.2in} (t\gg \beta\gg J^{-1})
\end{align}
From this, we would conclude that the scrambling time
\begin{align}
t_{s}\sim\beta
\end{align}
at low temperatures.  We note that power-law decay of the correlation function (\ref{eq:thermal_circle}), at sufficiently long times that are shorter than the inverse level spacing, has been observed for the SYK model \cite{Altland_SYK}, with the same exponent as in Eq. (\ref{eq:t6}).

We conclude this section by observing that the expression for the out-of-time-ordered commutator (\ref{eq:C_Complete}) dramatically simplifies in two particular cases.  First, if the operators initially have {disjoint} support, then we find that
\begin{align}
{\overline{C_{\beta}(t)}}{} = \frac{\Tr(V^{2}W^{2})}{N}\mathrm{Re}\Bigg[1 - \frac{\beta\,I_{1}(2\beta J + 2iJt)}{(\beta + it)I_{1}(2\beta J)}\left[\frac{I_{1}(2iJt)}{iJt}\right]^{3}\Bigg]\nonumber
\end{align} 
Alternatively, if the operators initially overlap, and satisfy $V\,W = 0$, then
\begin{align}\label{eq:C_Avg_VW0}
&{\overline{C_{\beta}(t)}}{} = \frac{\Tr(V^{2})\Tr(W^{2})}{N^{2}}\left[1 - \left(\frac{I_{1}(2iJt)}{iJt}\right)^{2}\right]
\end{align}
In the latter case, the out-of-time-ordered commutator has no temperature dependence to leading order in the large-$N$ limit.  An example is given by taking $V(0) = Z_{\rb} P^{(+)}_{\sB}$,  $W(0) = Z_{\rb'} P^{(-)}_{\sB}$, where $Z_{\rb}$ is a Pauli $Z$ operator at site $\rb$, while $P^{(\pm)}_{\rb} \equiv (1\pm Z_{\rb})/2$ is a projector onto $Z_{\rb}=\pm 1$.  While these operators commute and are each traceless, their product vanishes since $P^{(+)}_{\sB}P^{(-)}_{\sB} = 0$.

\subsection{Comparison with a Chaotic Quantum System} 
It is instructive to compare the behavior of the ensemble-averaged OTOC with the behavior of the same quantity in a ``chaotic" quantum system with $\mN$ degrees of freedom, in which the OTOC exhibits exponential growth in a parametrically large window of time
\begin{align}\label{eq:OTOC_chaos}
C_{\beta}(t)_{\mathrm{chaos}} \sim \frac{1}{\mN}e^{(2\pi t/\beta)\varkappa}   \hspace{.25in} (t_{r}\ll t\ll t_{s})
\end{align}
As indicated, this functional form is valid at times much larger than the \emph{relaxation time} of the system 
\begin{align}
t_{r}\sim \beta
\end{align}
which defines the timescale over which time-ordered correlation functions decay to their asymptotic values.  Furthermore, $t$ is much smaller than the \emph{scrambling time}
\begin{align}
 t_{s} \sim \frac{\beta}{2\pi\varkappa}\log\mN
 \end{align}
at which the OTOC reaches an $O(1)$ value. The parameter $\varkappa$ is the dimensionless Lyapunov exponent, which is known to satisfy the bound $0 \le \varkappa \le 1$ \cite{maldacena2016}.  In the SYK model of $2\mN$ Majorana fermions with all-to-all four-fermion interactions, for example, the OTOC takes the form (\ref{eq:OTOC_chaos}), with the exponent $\varkappa \rightarrow 1$ as $\beta\rightarrow\infty$.
 
We may draw two comparisons between a chaotic quantum system and the ensemble-averaged OTOC (\ref{eq:C_Complete}).  First, we observe that a typical Hamiltonian scrambles much more quickly than a chaotic quantum system with the same number of degrees of freedom, temperatures, and typical energy scales.  To draw this comparison, we fix $N \sim \exp(\mN)$, and require that the extensive energy scale $J$ in our random matrix model is fixed as $J \sim \mathcal{J}\mN$, where $\mathcal{J}$ is the typical strength of a local interaction in the chaotic quantum system.  In this case, at sufficiently low temperatures $\beta \mathcal{J} \gg \mN^{-1}$, the regime $\beta \ll t \ll (\beta/2\pi\varkappa)\log\mN$ over which $C_{\beta}(t)_{\mathrm{chaotic}}$ grows   exponentially (\ref{eq:OTOC_chaos}), corresponds to times that are much longer than the scrambling time of our model, $t_{s}\sim 1/J$, and in this regime, $\overline{C_{\beta}(t)} \sim O(1)$. 

We observe that a typical Hamiltonian scrambles faster than a chaotic quantum system since the typical range of an interaction will be of order $\mN$, unlike systems with local interactions or all-to-all $q$-body interactions, for which $q/\mN \rightarrow 0$ as $\mN\rightarrow\infty$, such as the SYK model. As we show in Sec. \ref{sec:symmetry}, the ``fast scrambling" property of our model is intrinsically linked to the invariance of our ensemble of Hamiltonians under unitary conjugation. 

We may draw a second comparison with a chaotic quantum system by identifying the relaxation and scrambling times for observables in our model, and extracting the behavior of $\overline{C_{\beta}(t)}$ in this intermediate regime.  For traceless operators $V$ and $W$, which are initially disjoint, we observe that the relaxation time is strictly zero in the large-$N$ limit, since ensemble-averaged time-ordered correlation functions such as 
\begin{align}
&\overline{\langle V(t)W(0) \rangle_{\beta}} = O(N^{-1})\\
&\overline{\langle W(0)V(t)^{2}W(0) \rangle_{\beta}} = \frac{\Tr(V^{2})\Tr(W^{2})}{N}
\end{align}
immediately reach their asymptotic values.  As a result, for these observables, there is a tunable separation between relaxation and scrambling times. In the large-$N$ limit, the growth of $\overline{C_{\beta}(t)}$ for disjoint operators in the window $0 \ll t \ll J^{-1}$ is given by
 \begin{align}
 \overline{C_{\beta}(t)} =\frac{\Tr(V^{2}W^{2})}{N}\left[J^{2}t^{2}\left( \frac{7}{2} - \frac{3}{\beta J}\frac{I_{2}(2\beta J)}{I_{1}(2\beta J)}\right) + O(J^{4}t^{4})\right]\nonumber
 \end{align}
The quantity in parentheses is an $O(1)$ constant for any value of $\beta J$.  For these observables, the early-time decay of the correlation function (\ref{eq:thermal_circle}) is also quadratic, i.e.
\begin{align}
\overline{\frac{F_{\beta}(t)}{F_{\beta}(0)}} = 1 - 8J^{2}t^{2}&\left[1 - \frac{6\,I_{2}(\beta J/2)}{\beta J \,I_{1}(\beta J /2)} - \frac{I_{2}(\beta J/2)^{2}}{I_{1}(\beta J /2)^{2}}\right]\nonumber\\ 
&+ O(J^{3}t^{3})\nonumber
\end{align}
when $Jt \ll 1$.  The quadratic time-dependence of out-of-time-ordered quantities within this window between typical relaxation and scrambling times contrasts with the exponential growth of the OTOC in a chaotic system.  

\section{Scrambling from Symmetry}\label{sec:symmetry}
While the calculations and results presented in the previous section are for Hamiltonians drawn from the Gaussian unitary ensemble in the large-$N$ limit, certain features of the out-of-time-ordered commutator are universal, when averaging over an ensemble of Hamiltonians that are related by unitary transformations
\begin{align}\label{eq:Unitary_Symmetry}
H \,\longrightarrow\,\mU H\mU^{\dagger} 
\end{align}
where $\mU$ is any $N\times N$ unitary matrix.  Given any Hamiltonian, the ensemble (\ref{eq:Unitary_Symmetry}) corresponds to a uniform distribution over all Hamiltonians with identical energy spectra.  This symmetry transformation is explicitly manifest for the ensemble of Hamiltonians drawn from distributions of the form $P(H) \sim \exp[-N\sum_{m}c_{m}\Tr(H^{m})]$, such as the Gaussian unitary ensemble. 

In the the Supplemental Material \cite{SM}, we demonstrate that any ensemble of Hamiltonians that is invariant under the transformation (\ref{eq:Unitary_Symmetry}) exhibits a power-law decay of the out-of-time-ordered commutator to its asymptotic value at late times, in the large-$N$ limit.  Without loss of generality, we state our result for the case where the density of states $\rho(\epsilon)$ is smooth, and non-zero only within a window of energies $[-2J, 2J]$; then, we show that the scrambling time
\begin{align}\label{eq:Scrambling_Conjecture}
t_{s} \sim J^{-1}
\end{align}
and that the OTOC decays to its asymptotic value as a power law that is at least as fast as
\begin{align}\label{eq:C_Unitary_Bound}
&\overline{C_{\beta}(\infty)} - \overline{C_{\beta}(t)}\,\,{\lesssim} \,\,t^{-2}\,\rho\,'(2J - t^{-1})
\end{align}
with $\rho'(\epsilon)$ the derivative of the density of states.  This bound is valid at any temperature, and for any choice of operators appearing in the OTOC, though the exact decay of the OTOC will depend on these parameters. As an example, for the Wigner semicircle distribution, there is a square-root singularity at the edge of the eigenvalue spectrum, so that $\rho\,'(2J - t^{-1})/t^{2} \sim (Jt)^{-3/2}$.  { If the derivative of the density of states is smooth, then (\ref{eq:C_Unitary_Bound}) predicts that
\begin{align}\label{eq:C_Unitary_Bound_2}
&\overline{C_{\beta}(\infty)} - \overline{C_{\beta}(t)}\,\,{\sim} \,\,(Jt)^{-\alpha}  \hspace{.2in} (Jt \gg 1)
\end{align}
with the exponent $\alpha \ge 2$, to leading order in $1/(Jt)$.  More generally, if the density of states is not a smooth function, the above expressions are modified by replacing $2J$ in (\ref{eq:Scrambling_Conjecture}) and (\ref{eq:C_Unitary_Bound}) with the difference in energy between the point at which the derivative of the density of states is most singular, and the center of the density of states. 

We believe that the power-law decay of the OTOC at times larger than the scrambling time, but smaller than the inverse level spacing ($t_{s} \ll t \ll N/J$), is generic for any unitary evolution with energy conservation.  Given a many-body Hamiltonian ($H$) with local interactions, a typical Hamiltonian drawn from the ensemble obtained by unitary conjugation (\ref{eq:Unitary_Symmetry}) will be highly non-local, so that the early-time growth of the averaged OTOC will not resemble the growth of the OTOC for the Hamiltonian of interest. The scrambling time $t_{s}$ sets a scale over which the OTOC varies appreciably, and effectively encodes the locality of $H$. From the expressions derived for the OTOC, we believe that at times of order the scrambling time, the locality of the Hamiltonian only enters through this timescale, so that the late-time behavior of the OTOC should be given by $\overline{C_{\beta}(\infty)} - \overline{C_{\beta}(t)} \sim (t_{s}/t)^{\alpha}$, with $\alpha \ge 2$, from Eq. (\ref{eq:C_Unitary_Bound}).  Indeed, power-law decay of this form has been seen in a range of systems including the SYK model \cite{Altland_SYK, Altland_SYK_2}, and in the many-body localized phase \cite{Chen2016}, though in the latter case, the numerically observed behavior appears to violate the condition $\alpha \ge 2$.

We conclude by reviewing the argument in the Supplemental Material \cite{SM}, that leads to the conditions (\ref{eq:Scrambling_Conjecture}) and (\ref{eq:C_Unitary_Bound}).  Our reasoning proceeds by observing that the ensemble-averaged OTOC may be calculated by first performing the unitary transformation $H\rightarrow \mU H \mU^{\dagger}$, and averaging over the choice of random unitary matrix $\mU$.  By studying the structure of the unitary average, we show that the decay of the out-of-time-ordered commutator to its asymptotic value is set by a power-law that depends on the singular structure of the density of states. 
We further verify our argument by calculating the OTOC for the non-Gaussian random matrix ensemble, defined by the probability distribution
\begin{align}
P(H) \sim \exp\left\{-{N}\left[\frac{1}{2}\Tr(H^{2}) - \frac{g}{4}\Tr(H^{4})\right]\right\}
\end{align}
in the Supplemental Material \cite{SM}.  For this ensemble, the exact eigenvalue distribution is known in the large-$N$ limit \cite{Non_Gaussian}, and the distribution is non-zero over an interval of width $\sim O(g^{-1/4})$, when $g \gg 1$.  We verify from the exact calculation that the scrambling time indeed grows as $t_{s} \sim g^{1/4}$ and that the power-law approach of the out-of-time ordered correlator to its asymptotic value satisfies the bound (\ref{eq:C_Unitary_Bound}).

\section{Coherent Scrambling}
The growth of out-of-time ordered correlation functions describes the spreading of local perturbations under Heisenberg evolution, and the delocalization of quantum information.  For systems with a holographic dual, certain out-of-time ordered correlation functions can be recast as \emph{time-ordered} {correlations} in the appropriate ``bulk" description.  This suggests that in a generic many-body system, the growth of out-of-time-ordered correlations can probe the presence of a dual description of the system of interest, if the out-of-time-ordered correlation function is viewed as the matrix element of an appropriately defined scattering matrix.

A scattering matrix with certain matrix elements corresponding to out-of-time-ordered correlation functions was introduced in Ref. \cite{Kitaev_Suh, Kitaev_IAS}, and the unitarity of this scattering matrix over a parametrically large window of time -- termed ``coherent scrambling" -- appears to be a feature of maximally chaotic quantum systems \cite{Kitaev_Suh}.  In the formulation of Ref. \cite{Kitaev_Suh, Kitaev_IAS}, however, the scattering matrix defines an ``embedding" of the Hilbert space of the system of interest into the Hilbert space of an auxiliary quantum system, chosen such that certain time-ordered correlation functions are approximately the same between the two.  

In this section, we take a brief detour from our previous discussion to present a formulation of the ideas presented in Ref. \cite{Kitaev_Suh, Kitaev_IAS}, by defining a scattering matrix that makes no reference to such an embedding in an auxiliary quantum system, which permits us to define ``coherent scrambling" for an arbitrary quantum many-body system.  Demanding that this scattering matrix is unitary naturally leads to the condition on certain out-of-time-ordered correlation functions, as shown in Eq. (\ref{eq:unitarity_condition}).  We note that the essential ideas presented here are found in Ref. \cite{Kitaev_Suh, Kitaev_IAS}.  We then demonstrate that this condition is satisfied by the out-of-time-ordered correlation functions computed for our random matrix problem over a small window of time when $Jt \ll 1$.  We note that the conditions derived here are necessary, but not sufficient, for the scattering matrix to be unitary.       

\subsection{Defining Coherent Scrambling in a Generic Many-Body System }

We begin by defining ``coherent scrambling" for a generic quantum many-body system. Consider a Hamiltonian $H$ for our system of interest, and let $V$ and $W$ be Hermitian operators that initially commute.  We assume, for the entirety of our discussion, that there is a tunable separation between relaxation ($t_{r}$) and scrambling times ($t_{s}$), and that we are probing the behavior of our system in this window $t_{r} \ll t \ll t_{s}$.  We now define the state 
\begin{align}\label{eq:TFD_H}
\ket{W(0)V(t)} \equiv \sum_{m}\frac{e^{-\beta E_{m}/2}}{\sqrt{N_{V,W}\, Z}}W(0)V(t)\ket{m}\otimes\ket{m}
\end{align}
with $V(t) = e^{-iHt}V(0)e^{iHt}$, and with $\ket{m}$, an eigenstate of $H$ with energy $E_{m}$.  That is, $\ket{W(0)V(t)}$ is the state defined by acting with the indicated operators on one copy of the thermofield double state of the system.  The  factors $Z \equiv \Tr(e^{-\beta H})$ and $N_{V,W} \equiv \langle V(t)W(0)^{2}V(t) \rangle_{\beta}$ guarantee that the state is normalized $\braket{W(0)V(t)\,|\,W(0)V(t)} = 1$.  

We now \emph{define} a scattering matrix $\mS$ that acts within the Hilbert space of states of the form (\ref{eq:TFD_H}), by the condition that its diagonal matrix elements are proportional to an out-of-time-ordered correlation function of the operators that define the state.  Specifically,
\begin{align}
\braket{W(0)V(t)|\mS|W(0)V(t)} \equiv \frac{\langle V(t)W(0)V(t)W(0)\rangle_{\beta}}{\langle W(0)V(t)^{2}W(0)\rangle_{\beta}}\nonumber
\end{align}
We may rewrite the above condition in a more suggestive form.  For notational convenience, we denote $\braket{W(0)V(t)|\mS|W(0)V(t)}$ simply as $\braket{\,\mS\,}$, and write
\begin{align}\label{eq:S_C_D}
\left\langle\,\mS\,\right\rangle = 1 + \frac{D_{\beta}(t) - C_{\beta}(t)}{\langle W(0)V(t)^{2}W(0)\rangle_{\beta}}
\end{align}
where $C_{\beta}(t) = (-1/2) \langle [V(t), W(0)]^{2}\rangle_{\beta}$ is the out-of-time-ordered commutator, as defined in (\ref{eq:C}), while
\begin{align}\label{eq:D}
&D_{\beta}(t) \equiv \frac{1}{2}\Big\langle \left[V(t), W(0)\right] \,\left\{V(t), W(0)\right\}\Big\rangle_{\beta}
\end{align}
with $\{\cdots\}$ denoting the anti-commutator. The matrix $\left\langle\,\mS\,\right\rangle\approx 1$ for times close to the relaxation time $t\rightarrow t_{r}$, when the out-of-time-ordered quantities $C_{\beta}$ and $D_{\beta}$ are small.  In a chaotic quantum system, the deviation from unity will be suppressed by $O(1/\mN)$.  This is consistent with the fact that a scattering matrix should act as the identity at sufficiently early times.

At times $t\gtrsim t_{r}$, when time-ordered correlation functions have decayed to near their asymptotic values, we assume that
\begin{align}\label{eq:assumption}
\langle W(0)V(t)^{2}W(0)\rangle_{\beta}\approx\langle V(t)W(0)^{2}V(t)\rangle_{\beta}
\end{align}
We note that $C_{\beta}(t)$ is real, and with the above assumption, $D_{\beta}(t)$ is purely imaginary when $t\gtrsim t_{r}$.  If we express the scattering matrix as $\mS = \exp\left(i \varepsilon \mM\right)$, then when the out-of-time-ordered correlation functions $|D_{\beta}(t)|$, $C_{\beta}(t)\ll 1$, we may expand the scattering matrix as $\mS = 1 + i\varepsilon \mM - (\varepsilon^{2}/2)\mM^{2} + \cdots$.  If $\mS$ is unitary, then $\mM$ is Hermitian, and we are forced to identify $D_{\beta}(t) \approx i\epsilon\langle\mM\rangle$ and $C_{\beta}(t)\approx (\epsilon^{2}/2)\langle\mM^{2}\rangle$, so that  
\begin{align}\label{eq:unitarity_condition}
|D_{\beta}(t)| \gg C_{\beta}(t)\hspace{.25in} (t\gtrsim t_{r})
\end{align}
when $|D_{\beta}(t)|$, $C_{\beta}(t)\ll 1$.  This condition can only be satisfied at sufficiently low temperatures, since $D_{\beta}(t)$ is exactly zero for any system at infinite temperature.

A more precise condition for the unitarity of the scattering matrix may be derived, if we make the additional assumption that $\max_{V, W}\langle \mM \rangle \sim O(||\mM||)$, i.e that for a particular choice of operators $V$ and $W$, the expectation value $\langle \mM \rangle$ is of the same order as the operator norm of $\mM$; this is a reasonable assumption, since $V$ and $W$ can be arbitrary Hermitian operators.  With this, we observe that the unitarity of the scattering matrix at times when $|D_{\beta}(t)|$, $C_{\beta}(t)\ll 1$ requires that for \emph{some choice} of operators $V$ and $W$, 
\begin{align}\label{eq:unitarity_condition_2}
C_{\beta}(t) \sim \frac{|D_{\beta}(t)|^{2}}{\langle W(0) V(t)^{2}W(0)\rangle_{\beta}} 
\end{align}
We define the \emph{coherent scrambling time} $t_{c}$ as the timescale over which Eq. (\ref{eq:unitarity_condition_2}) is satisfied; this defines a time over which the $\mS$ matrix becomes appreciably non-unitary. 
We note that the above condition need not be satisfied by the out-of-time-ordered correlation functions for an arbitrary choice of operators $V$ and $W$.  

For the SYK model of $2\mN$ Majorana fermions interacting via four-fermion interactions, and with typical interaction strength $\mJ$, the behavior of these out-of-time-ordered correlation functions for fermion operators\footnote{The OTOC for fermion operators is defined as in Eq. (\ref{eq:C}), with the commutator replaced by the anti-commutator, while the definition of $D_{\beta}(t)$ is the same as in Eq. (\ref{eq:D}). With this re-definition of the OTOC, our definition of the scattering matrix, and the condition for unitarity at early times presented in Eq. (\ref{eq:unitarity_condition_2}), are unchanged.} at low temperatures $1 \ll \beta\mJ\ll \mN$ is given by \cite{Kitaev_Suh, Yingfei_Discussion}
\begin{align}\label{eq:SYK_OTOC}
&C_{\beta}(t) \sim \frac{1}{\mathcal{N}}e^{2\pi t\varkappa/\beta}\hspace{.3in}
|D_{\beta}(t)| \sim \frac{\beta\mathcal{J}}{\mathcal{N}}e^{2\pi t\varkappa/\beta}
\end{align}
at times $\beta \ll t\ll O(\beta\log\mN)$.  As a result, the condition $|D_{\beta}(t)| \gg C_{\beta}(t)$ is satisfied throughout this window of time.  However, the second condition (\ref{eq:unitarity_condition_2}) appears to be violated by out-of-time-ordered correlation functions of the Majorana fermions. 

\subsection{Coherent Scrambling for a GUE Hamiltonian}
We now demonstrate that for a random, GUE Hamiltonian, the coherent scrambling condition (\ref{eq:unitarity_condition_2}) is satisfied over a short window of time, when $t \ll J^{-1}$.  First, for a GUE Hamiltonian, we determine that for traceless operators $V$ and $W$ that initially commute, the ensemble-averaged quantity $\overline{D_{\beta}(t)}$ is given by the following expression in the large-$N$ limit \cite{SM}:  
\begin{widetext}
\begin{align}
\overline{D_{\beta}(t)} = \,\,&i\frac{\Tr\left(VW\right)^{2}}{N^{2}}\frac{\beta }{I_{1}(2\beta J)} \frac{I_{1}(2iJt)}{it}\,\,\mathrm{Im}\left\{\frac{I_{1}(2iJt)}{iJt} F(\beta J + iJt, iJt) + \frac{I_{1}(2\beta J + 2iJt)}{\beta J + iJt}F(iJt, iJt)\right\}\\
&+i\frac{\Tr\left(V^{2}W^{2}\right)}{N}\,\,\mathrm{Im}\left\{\frac{\beta}{\beta + it}\frac{I_{1}(2\beta J + 2iJt)}{I_{1}(2\beta J)}\right\}\left[\frac{I_{1}(2iJt)}{iJt}\right]^{3} + O(N^{-1})\nonumber
\end{align}
\end{widetext}
As with the out-of-time-ordered commutator, we observe that $\overline{D_{\beta}(t)}$ strictly vanishes at a time $t = t_{0} \approx 1.916/J$. 


To understand whether a random Hamiltonian satisfies our ``coherent scrambling" condition, we note that for initially overlapping operators $V$ and $W$, whose product vanishes,  $\overline{D_{\beta}(t)}$ is trivially zero in the large-$N$ limit
\begin{align}
&{\overline{D_{\beta}(t)}} = 0 + O(N^{-1})\hspace{.2in} (V\,W = 0)
\end{align}
For the three other classes of operators, $\overline{D_{\beta}(t)}$ grows linearly at early times.  Specifically, for disjoint, traceless operators $V$ and $W$, for which there is a tunable separation between relaxation and scrambling times, we observe that in the large-$N$ limit
\begin{align}
\overline{D_{\beta}(t)} = \frac{\Tr(V^{2}W^{2})}{N}\left[iJt\left(\frac{2\,I_{2}(2\beta J)}{I_{1}(2\beta J)}\right) + O(Jt)^{3}\right]
\end{align} 
when $0 \ll t\ll J^{-1}$.  
This immediately implies that the ensemble-averaged out-of-time-ordered correlation functions satisfy the condition (\ref{eq:unitarity_condition_2}) at low temperatures
\begin{align}
\overline{C_{\beta}(t)} \sim \overline{\frac{|D_{\beta}(t)|^{2}}{\langle W(0) V(t)^{2}W(0)\rangle_{\beta}}} \,\,\,\,\,\,\,(\beta J \gg 1)
\end{align}
and at times $0 \ll t \ll J^{-1}$.  In other words, at low temperatures, the coherent scrambling time -- the regime where the above condition is satisfied -- is of the same order as the scrambling time
\begin{align}
t_{c}\sim t_{s} \sim J^{-1}
\end{align} 
The condition (\ref{eq:unitarity_condition_2}) does not appear to be satisfied by the the out-of-time-ordered correlation functions of the Majorana fermions in the SYK model at low temperatures, as can be seen from Eq. (\ref{eq:SYK_OTOC}); we are unaware if this condition is satisfied by other operators in the theory.

 \acknowledgments
SV is supported by the DOE Office of Basic Energy Sciences,  Division  of  Materials Sciences  and   Engineering under Award de-sc0010526.  AV was supported by a Simons Investigator award. We thank  Yi-Zhuang You and Yingfei Gu for helpful discussions and  for bringing Ref. \cite{Kitaev_IAS} to our attention.

\appendix
\section{Details of the OTOC Calculation for Initially Disjoint Operators}
We outline the calculation of the ensemble-averaged out-of-time-ordered correlator
\begin{align}
\overline{\left\langle V(t)W(0)V(t)W(0) \right\rangle_{\beta}} 
\end{align}
when the Hermitian operators $V(0)$ and $W(0)$ are disjoint (i.e. their support is non-overlapping), so that $\Tr(VW) = N^{-1}\Tr(V)\Tr(W)$. We first calculate the above quantity when $V$ and $W$ are traceless.  The generalization to operators with a non-zero trance is straightforward, and is presented at the end of this section.

We wish to perform an average over the Gaussian unitary ensemble for the Hamiltonian $H$, i.e. $H$ is distributed according to
\begin{align}
P(H) \sim \exp\left[-\frac{N}{2}\Tr(H^{2})\right].
\end{align}
We note that this corresponds to setting $J = 1$ in the distribution quoted in the main text.  We restore the dependence on $J$ at the end of our calculation. In the large-$N$ ($N\rightarrow\infty$) limit, only non-crossing contractions of $H$ contribute to leading order in $N$.  This means, for example, that
$\overline{H^{4}} = \contraction{}{H}{}{H}HH\contraction{}{H}{}{H}HH + \contraction[2ex]{}{H}{HH}{H}\contraction{H}{H}{}{H}HHHH + O(N^{-1}) = 2\,\boldsymbol{1}_{N\times N} + O(N^{-1})$.  For a traceless operator $\mathcal{O}$, we observe that in the large-$N$ limit, $\overline{H^{n}\mathcal{O}H^{m}} = \overline{H^{n}} \,\mathcal{O}\,\overline{H^{m}}$ since all other terms that contract $H$ across the operator $\mathcal{O}$ are proportional to its trace, and therefore vanish.
As a consequence, we observe that in the large-$N$ limit,
$\overline{H^{n}\,V\,H^{m}\,W\,H^{p}\,V\,H^{q}\,W} =\overline{H^{n}}\,V\,\overline{H^{m}}\,W\,\overline{H^{p}}\,V\,\overline{H^{q}}\,W$
when $V$, $W$ are traceless. Using these properties, we now compute the averaged OTOC that appears in the expression for $C_{\beta}(t)$.  First, from the fact that only non-crossing contractions contribute to the GUE average at leading order in $N$, we observe that the ensemble-average of the products of traces of operators, factorizes in the large-$N$ limit.  Therefore, 
\begin{align}\label{eq:OTOC_Avg}
&\overline{\left\langle V(t)W(0)V(t)W(0) \right\rangle_{\beta}}
 = \frac{\overline{\,\,\Tr\left[e^{-\beta H} V(t)W(0)V(t)W(0) \right]\,}}{\overline{\,\,\Tr\left[e^{-\beta H}\right]^{\,^{}}}\,\,\,} 
\end{align}
where we have used the simplification in the large-$N$ limit to rewrite the third line, leading to the final expression.  Letting $M_{\beta}(t) \equiv \overline{\,\,\Tr\left[e^{-\beta H} V(t)W(0)V(t)W(0) \right]\,}$, we now observe that 
\begin{align}\label{eq:M}
M_{\beta}(t) = \sum_{n,m,p,q = 0}^{\infty}&\frac{(-\beta-it)^{n}(it)^{m}(-it)^{p}(it)^{q}}{n! \,m! \,p! \,q!} \\
&\times\overline{\,\,\Tr\left[H^{n}\,V\,H^{m}\,W\,H^{p}\,VH^{q}\,W\right]\,\,}\nonumber
\end{align}
We may evaluate this expression, by observing that $\overline{H^{n}} = N^{-1}\Tr\left(\,\overline{H^{n}}\,\right)\,\boldsymbol{1}_{N\times N} = \overline{\,\lambda^{n}\,}\boldsymbol{1}_{N\times N}$ where
\begin{align}\label{eq:Wigner_Avg}
\overline{\,\lambda^{n}\,} = \frac{1}{2\pi}\int_{-2}^{2}d\lambda\,\lambda^{n}\sqrt{4-\lambda^{2}}
\end{align}
is the $n$th moment of the Wigner semi-circle distribution, which describes the probability distribution of a single eigenvalue of the matrix $H$.  Clearly $\overline{\,\lambda^{n}\,} = 0$ for odd $n$.  When $n$ is even, we find that
\begin{align}
\overline{\,\lambda^{2n}\,} = \frac{(2n)!}{(n!)^{2}\,(n+1)}
\end{align}
We simplify the expression (\ref{eq:M}) to obtain
\begin{widetext}
\begin{align}\label{eq:M}
M_{\beta}(t) = \sum_{n,m,p,q = 0}^{\infty}&\frac{(-\beta-it)^{n}(it)^{m}(-it)^{p}(it)^{q}}{n! \,m! \,p! \,q!} \overline{\lambda^{n}}\cdot\overline{\lambda^{m}}\cdot\overline{\lambda^{p}}\cdot\overline{\lambda^{q}} \,\Tr\left(V^{2}W^{2}\right) = \Tr\left(V^{2}W^{2}\right)\,f(it)^{2}\,f(-\beta - it)\,f(-it)
\end{align}
\end{widetext}
where
\begin{align}
f(x) \equiv \sum_{n=0}^{\infty} \frac{x^{n}\,\overline{\lambda^{n}}}{n!} =  \sum_{n=0}^{\infty} \frac{x^{2n}}{(n!)^{2}\,(n+1)} = \frac{I_{1}(2x)}{x}
\end{align}
where $I_{1}(x)$ is the modified Bessel function of the first kind.  We observe that $f(x) = f(-x)$, so that \begin{align}\label{eq:M}
M_{\beta}(t) = {\Tr\left(V^{2}W^{2}\right)}\,\,\frac{I_{1}(2\beta + 2it)}{\beta + it}\left(\frac{I_{1}(2it)}{it}\right)^{3}
\end{align}
Finally, we observe that $\overline{\Tr(e^{-\beta H})} = N\,f(-\beta)$.   

We now use the above expressions to determine the ensemble-averaged out-of-time-ordered commutator (OTOC) between operators $V$ and $W$, that are initially disjoint.  Expanding the definition of the OTOC, we observe that
\begin{align}\label{eq:OTOC_Trace}
C_{\beta}(t) &= \frac{1}{2}\langle V(t) W(0)^{2}V(t)\rangle_{\beta} + \frac{1}{2}\langle W(0)V(t) ^{2}W(0)\rangle_{\beta}\nonumber\\
&- \mathrm{Re}\Big[ \langle V(t) W(0) V(t) W(0)\rangle_{\beta} \Big]
\end{align}
The ensemble-averaged OTOC may be evaluated, using the properties derived previously.  To leading order in the large-$N$ expansion, the first two terms in (\ref{eq:OTOC_Trace}) are equivalent after averaging:
\begin{align}
&\overline{\langle V(t) W(0)^{2}V(t)\rangle_{\beta}} = \overline{\langle W(0)V(t) ^{2}W(0)\rangle_{\beta}} \nonumber
\end{align}
and that
\begin{align}
\overline{\langle V(t) W(0)^{2}V(t)\rangle_{\beta}}={N^{-2}}{\Tr\left( V^{2}\right) \Tr\left( W^{2}\right)}\label{eq:first_terms_OTOC}
\end{align}
Here, we have used the fact that $V(0)$ and $W(0)$ are disjoint to simplify the expression.  Using these relations, as well as the expression (\ref{eq:M}), we conclude that when $V(0)$ and $W(0)$ are disjoint, the OTOC is given by the expression
\begin{align}
\overline{C_{\beta}(t)} = \frac{\Tr(V^{2})\Tr(W^{2})}{N^{2}}\mathrm{Re}\left [1 - \frac{\beta I_{1}(2\beta + 2it)}{(\beta + it)\,I_{1}(2\beta)}\left(\frac{I_{1}(2it)}{it}\right)^{3}\right]\nonumber
\end{align}
as presented in the main text.  

When $V$ and $W$ are not traceless, only the overall prefactor in the above expression is different.  This can be easily seen by rewriting $V(t)$ and $W$ as
\begin{align}
V(t) &= \mathcal{O}_{V}(t) + \frac{1}{N}\Tr\left(V\right)\nonumber\\
W(0) &= \mathcal{O}_{W} + \frac{1}{N}\Tr\left(W\right)\nonumber
\end{align}
where $\mathcal{O}_{V}$ and $\mathcal{O}_{W}$ are traceless, Hermitian operators.  We then find that
\begin{widetext}
\begin{align}
\overline{ \langle V(t) W(0) V(t) W(0)\rangle_{\beta}} = \overline{\langle \mathcal{O}_{V}(t)\mathcal{O}_{W}\mathcal{O}_{V}(t)\mathcal{O}_{W}\rangle_{\beta}} + \frac{1}{N^{2}}\Tr\left(W\right)^{2}\overline {\, \langle\mathcal{O}_{V}(t)^{2}\,\rangle_{\beta}} + \frac{1}{N^{2}}\Tr\left(V\right)^{2} \overline{\, \langle\mathcal{O}_{W}(0)^{2}\,\rangle_{\beta}} + \frac{1}{N^{4}}\Tr\left(V\right)^{2}\Tr\left(W\right)^{2}\nonumber
\end{align} 
\end{widetext}
We now observe that
\begin{align}
\overline{\,\langle\mathcal{O}_{V}^{\,2}\rangle_{\beta}\,} = \frac{1}{N}\left(\Tr\left(V^{2}\right) - \frac{1}{N}\Tr\left(V\right)^{2}\right)
\end{align}
and similarly for $\overline{\,\langle\mathcal{O}_{W}^{\,2}\rangle_{\beta}\,} $.  Using (\ref{eq:M}), along with the above expression, we may write the ensemble-averaged OTOC after defining the function
\begin{align}
g(V,W) = &\frac{1}{N^{2}}\Tr\left(V^{2}\right)\Tr\left(W^{2}\right) - \frac{1}{N^{3}}\Tr\left(V^{2}\right)\Tr\left(W\right)^{2}\nonumber\\
&- \frac{1}{N^{3}}\Tr\left(V\right)^{2}\Tr\left(W^{2}\right) + \frac{1}{N^{4}}\Tr\left(V\right)^{2}\Tr\left(W\right)^{2}\nonumber
\end{align}
This leads to the expression 
\begin{align}
\overline{C_{\beta}(t)} = g(V,W)\, \mathrm{Re}\left [1 - \frac{\beta I_{1}(2\beta + 2it)}{(\beta + it)\,I_{1}(2\beta)}\left(\frac{I_{1}(2it)}{it}\right)^{3}\right]\nonumber
\end{align}
A similar calculation leads to the expression for $\overline{D_{\beta}(t)}$ presented in the main text.

\subsection{Asymptotic Behavior of $\overline{C_{\beta}(t)}$ and $\overline{D_{\beta}(t)}$ for Initially Disjoint Operators}
Consider the expression for $\overline{C_{\beta}(t)}$, for traceless, Hermitian operators $V$ and $W$, that are initially \emph{disjoint}.  Without loss of generality, we take these to be simple Pauli operators that satisfy $V^{2} = W^{2} = 1$.  In this case we observe that at early times
\begin{align}
&\lim_{t\rightarrow 0}\overline{C_{\beta}(t)} = \left(\frac{7}{2} - \frac{3\,I_{2}(2\beta)}{\beta\,I_{1}(2\beta)}\right)t^{2} + O(t^{3})\\
&\lim_{t\rightarrow 0}\overline{D_{\beta}(t)} = i\frac{2\,I_{2}(2\beta)}{I_{1}(2\beta)}t + O(t^{3})
\end{align}


At high temperatures  ($\beta \rightarrow 0$), we observe that 
\begin{align}
&\lim_{\beta\rightarrow 0}\overline{C_{\beta}(t)} = 1 - \left(\frac{I_{1}(2it)}{it}\right)^{4} + O(\beta^{2})\label{eq:C_Inf}\\
&\lim_{\beta\rightarrow 0}\overline{D_{\beta}(t)} = \frac{2\beta}{it} I_{2}(2it)\left(\frac{I_{1}(2it)}{it}\right)^{3} + O(\beta^{2})\label{eq:D_Inf}
\end{align}
Finally, at low temperatures ($\beta \rightarrow \infty$), 
\begin{align}
\lim_{\beta\rightarrow\infty}\overline{C_{\beta}(t)} &= 1 - \left(\frac{I_{1}(2it)}{it}\right)^{3}\left[\cos(2t)
 + \frac{3t}{2\beta}\sin(2t)
 \right] + O(\beta^{-2})\nonumber\\
\lim_{\beta\rightarrow\infty}\overline{D_{\beta}(t)}  &= i\left(\frac{I_{1}(2it)}{it}\right)^{3}\left [\sin(2t) - \frac{3t}{2\beta}\cos(2t)\right] + O(\beta^{-2})\nonumber\\
\end{align}

\section{Details of the OTOC Calculation for Initially Overlapping Operators}
We now outline the calculation of the OTOC, when the operators $V(0)$ and $W(0)$ have overlapping support.  We assume, for simplicity, that $V$ and $W$ are traceless $\Tr(V) = \Tr(W) = 0.$
In this case, there are three different contributions to an ensemble-averaged quantity of the form
\begin{align}\label{eq:overlapping_op}
\overline{\Tr(H^{n}VH^{m}WH^{p}VH^{q}W)}
\end{align}
which are given below
\begin{align}
&\hspace{.0in}{\bf I.\hspace{.1in}} \Tr(\overline{H^{n}}\,V\,\overline{H^{m}}\,W\,\overline{H^{p}}\,V\,\overline{H^{q}}\,W)\nonumber\\
&\hspace{.0in}{\bf II.\hspace{.1in}}\overline{\displaystyle\sum_{r=0}^{n-1}\sum_{s=0}^{p-1}{\Tr\big(H^{r}\contraction{}{H}{H^{n-r-1}VH^{m}WH^{s}}{H}H H^{n-r-1}VH^{m}WH^{s}H H^{p-s-1}VH^{q}W\big)}}\nonumber\\
&\hspace{.0in}{\bf III.\hspace{.1in}}\overline{\displaystyle\sum_{r=0}^{m-1}\sum_{s=0}^{q-1}{\Tr\big(H^{n}VH^{r}\contraction{}{H}{H^{m-r-1}WH^{p}VH^{s}}{H}H H^{m-r-1}WH^{p}VH^{s}H H^{q-s-1}W\big)}}\nonumber
\end{align}

The second and third terms in the above expression vanish if $\Tr(VW) = 0$.  We now observe that the second term may be simplified as follows, in the large-$N$ limit
\begin{align}
&\frac{1}{N}\sum_{r=0}^{n-1}\sum_{s=0}^{p-1}\overline{\Tr\big(H^{p+r-s-1}VH^{q}W\big)}\hspace{.1in}\overline{\Tr\big(H^{n+s-r-1}VH^{m}W\big)}\nonumber\\
&=\frac{\Tr(VW)^{2}}{N}\overline{\lambda^{m}}\,\,\,\overline{\lambda^{q}}\,c_{n,p} + O(N^{0})
\end{align}
where
\begin{align}\label{eq:cnp}
c_{n,p} \equiv \sum_{r=0}^{n-1}\sum_{s=0}^{p-1}\overline{\lambda^{p+r-s-1}}\,\,\overline{\lambda^{n+s-r-1}}
\end{align}
and $\overline{\lambda^{n}}$ denotes an average with respect to the Wigner semicircle distribution
$\rho(\lambda) \equiv \frac{1}{2\pi}\sqrt{4-\lambda^{2}}\,\,\Theta(2 - |\lambda|)$. Using these expressions, we rewrite $c_{n,p}$ as
\begin{align}
c_{n,p} = \int dx\,dy\,\rho(x)\,\rho(y) \,\frac{x^{p}-y^{p}}{x-y}\,\frac{x^{n}-y^{n}}{x-y}
\end{align}
Similarly, the third contribution to (\ref{eq:overlapping_op}) evaluates to $N^{-1}\Tr(VW)^{2}\overline{\lambda^{n}}\,\,\overline{\lambda^{p}}\,c_{m,q}$.  

Using the above relations, and after defining the function $F(z,w)$, by the expression
\begin{align}
F(z,w) = \int dx \,dy \,\rho(x)\rho(y)\frac{(e^{-wx} - e^{-wy})(e^{-zx} - e^{-zy})}{(x-y)^{2}}
\end{align}
we evaluate the ensemble-averaged out-of-time-ordered correlator to obtain
\begin{widetext}
\begin{align}\label{eq:OTOC_overlapping}
&\overline{\langle V(t)W(0)V(t)W(0)\rangle_{\beta}}= \frac{\beta}{N\,I_{1}(2\beta)}\sum_{n,m,p,q = 0}^{\infty}\frac{(-\beta-it)^{n}(it)^{m}(-it)^{p}(it)^{q}}{n! \,m! \,p! \,q!} \overline{\,\,\Tr\left[H^{n}\,V\,H^{m}\,W\,H^{p}\,VH^{q}\,W\right]\,\,}\\
&= \frac{\Tr\left(VW\right)^{2}}{N^{2}}\frac{\beta}{I_{1}(2\beta)} \frac{I_{1}(2it)}{it}\left[\frac{I_{1}(2it)}{it}F(\beta + it, it) + \frac{I_{1}(2\beta + 2it)}{\beta + it}F(it, it)\right]
+\frac{\Tr\left(V^{2}W^{2}\right)}{N}\,\,\frac{\beta}{\beta + it}\frac{I_{1}(2\beta + 2it)}{I_{1}(2\beta)}\left(\frac{I_{1}(2it)}{it}\right)^{3}\nonumber
\end{align}
\end{widetext}
We may now obtain the final expression for the ensemble-averaged out-of-time-ordered commutator $\overline{C_{\beta}(t)}$.  We observe that to leading order in the large-$N$ limit, ensemble-averaged quantities $\overline{\langle V(t) W(0)^{2}V(t)\rangle_{\beta}}$ and $\overline{\langle W(0)V(t) ^{2}W(0)\rangle_{\beta}}$ are equal and given by  
\begin{align}\label{eq:avg_overlapping_1}
\overline{\langle V(t) W(0)^{2}V(t)\rangle_{\beta}} = \overline{\langle W(0)V(t) ^{2}W(0)\rangle_{\beta}}\nonumber
\end{align}
\begin{align}
\hspace{.1in}= \left(\frac{I_{1}(2it)}{it}\right)^{2}&\left[\frac{\Tr(V^{2}W^{2})}{N} - \frac{\Tr(V^{2})\Tr(W^{2})}{N^{2}}\right]\nonumber\\
&+  \frac{\Tr\left( V^{2}\right) \Tr\left( W^{2}\right)}{{N^{2}}}
\end{align}
which generalizes the expression in (\ref{eq:first_terms_OTOC}).  From this expression, and from (\ref{eq:OTOC_overlapping}), we obtain the expression for $\overline{C_{\beta}(t)}$ that appears in the main text.

From (\ref{eq:avg_overlapping_1}) and (\ref{eq:OTOC_overlapping}), we may construct the ensemble-averaged quantities $\overline{C_{\beta}(t)}$ and $\overline{D_{\beta}(t)}$.  For simplicity of presentation, however, we choose to restrict ourselves to the case where $V(0) = W(0) = X_{\rb}$ where $X_{\rb}$ is a Pauli operator, so that $X_{\rb}^{2} = 1$.  Then, the ensemble averaged quantities take the form
\begin{widetext}
\begin{align}
&\overline{C_{\beta}(t)} = 1 - \frac{\beta}{I_{1}(2\beta)} \frac{I_{1}(2it)}{it}\mathrm{Re}\left\{\frac{I_{1}(2it)}{it} F(\beta + it, it) + \frac{I_{1}(2\beta + 2it)}{\beta + it}F(it, it)\right\}
-\mathrm{Re}\left\{\frac{\beta}{\beta + it}\frac{I_{1}(2\beta + 2it)}{I_{1}(2\beta)}\right\}\left[\frac{I_{1}(2it)}{it}\right]^{3}\nonumber\\
&\overline{D_{\beta}(t)} = \frac{i\beta}{I_{1}(2\beta)} \frac{I_{1}(2it)}{it}\mathrm{Im}\left\{\frac{I_{1}(2it)}{it} F(\beta + it, it) + \frac{I_{1}(2\beta + 2it)}{\beta + it}F(it, it)\right\}
+i\mathrm{Im}\left\{\frac{\beta}{\beta + it}\frac{I_{1}(2\beta + 2it)}{I_{1}(2\beta)}\right\}\left[\frac{I_{1}(2it)}{it}\right]^{3}\nonumber
\end{align}
\end{widetext}

At infinite temperature, the out-of-time-ordered commutator takes the form
\begin{align}
\overline{C_{0}(t)} = 1 -  2\left(\frac{I_{1}(2it)}{it}\right)^{2}\mathrm{Re}\Big\{F(it, it) \Big\} - \left(\frac{I_{1}(2it)}{it}\right)^{4}\nonumber
\end{align}
Using the early-time, high-temperature behavior of $F(\beta + it, it) = -t^{2} + i\beta t + O(\beta^{2}t^{2})$, and that $I_{1}(2\epsilon)/\epsilon = 1+(\epsilon^{2}/2) + O(\epsilon^{4})$ as $\epsilon\rightarrow 0$, we find that the early-time behavior of the out-of-time-ordered quantities as $\beta \rightarrow 0$ is
\begin{align}
&{\overline{C_{\beta}(t)} = 4t^{2} + O(\beta^{2}t^{2})}\hspace{.2in}(\beta,\,t\rightarrow 0)\\
&{\overline{D_{\beta}(t)} = 2i\beta t + O(\beta^{2}t^{2})}
\end{align}

From the asymptotic form of $F(\beta + it, it)$ at long times, and using the fact that $I_{1}(2it)/it \sim \cos\left[\frac{\pi}{4} + 2t\right]/t^{3/2}$ as $t\rightarrow\infty$, and $\beta \rightarrow 0$, we find that
 \begin{align}
&{ \overline{C_{\beta}(t)} \sim 1 - \frac{\cos^{2}\left[\frac{\pi}{4} + 2t\right]}{t^{4}} + O(t^{-4}\beta, t^{-5})}
\end{align}

At low temperatures, we derive the asymptotic behavior of the out-of-time-ordered correlation functions by observing that as $\beta \rightarrow \infty$, 
\begin{align}
&\frac{\beta}{I_{1}(2\beta)} = e^{-2\beta} 2\sqrt{\pi}\,\left[\beta^{3/2} + O(\beta^{5/2})\right]\nonumber\\
&\frac{\beta}{\beta + it}\frac{I_{1}(2\beta + 2it)}{I_{1}(2\beta)} = e^{2it} + O(\beta^{-1})
\end{align}
Using these facts, and with the asymptotic behavior of $F(\beta + it, it)$ low temperatures, we determine that the OTOC grows quadratically $\sim J^{2}t^{2}$ in time at early times, while $D_{\beta}(t)\sim iJt$. 

\begin{figure}[t]
 \includegraphics[trim = 0 0 0 0, clip = true, width=0.25\textwidth, angle = 0.]{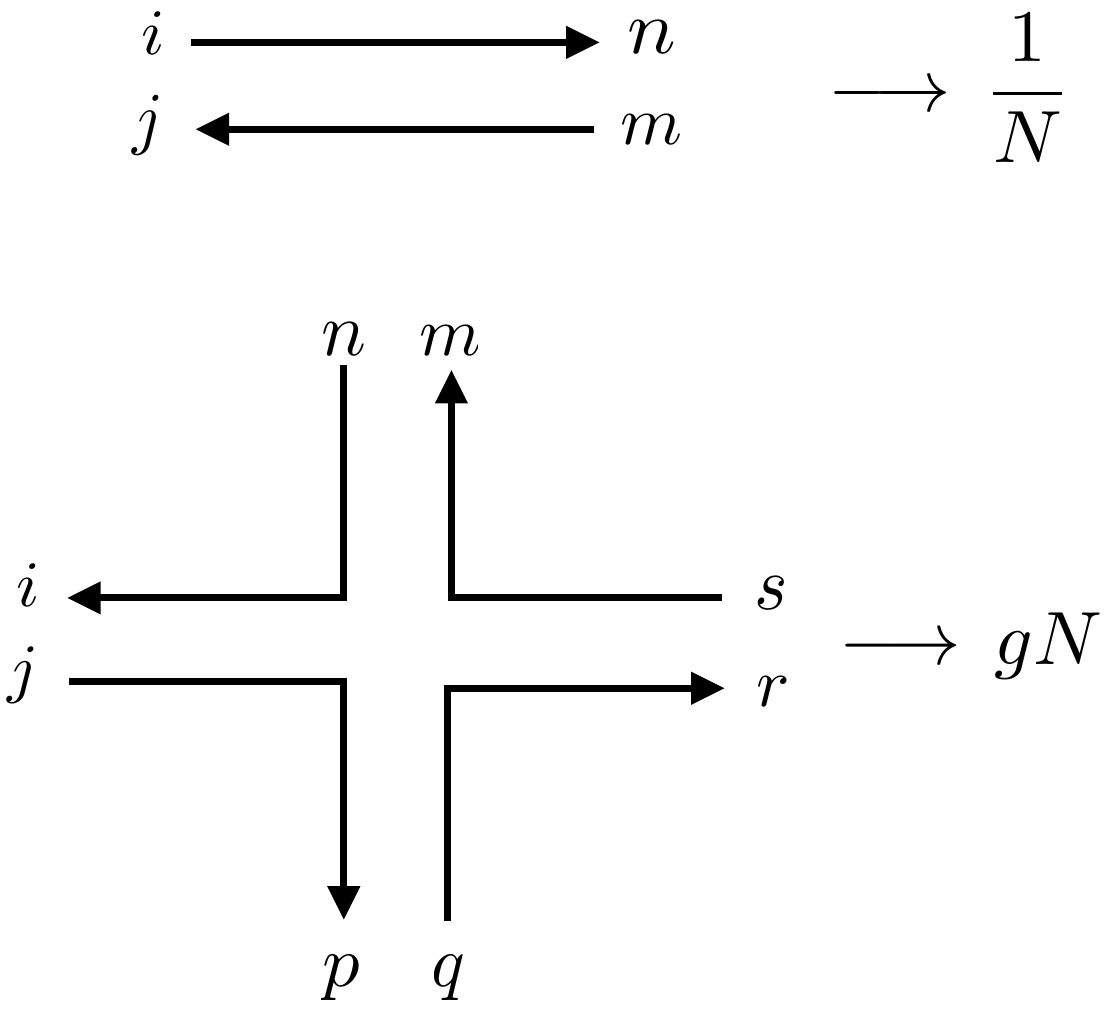}
 \caption{The propagator and four-point vertex appearing in the diagrammatic calculation of correlation functions for the non-Gaussian distribution in (\ref{eq:NG}).}
  \label{fig:Feyn_Diag}
\end{figure}

\section{Non-Gaussian Random Matrix Ensembles}
We may also consider the non-Gaussian random matrix ensemble, with probability density
\begin{align}\label{eq:NG}
P(H) \sim \exp\left\{-{N}\left[\frac{1}{2}\Tr(H^{2}) - \frac{g}{4}\Tr(H^{4})\right]\right\}
\end{align}
Correlation functions with respect to this distribution may be evaluated, by employing a diagrammatic expansion in powers of $g$, using the Feynman diagrams shown in Fig. \ref{fig:Feyn_Diag}.  Let the Hermitian operators $V$ and $W$ be traceless, and have non-overlapping support at time $t = 0$.  In this case, the correlation functions of the form $\overline{\Tr(H^{n}VH^{m}WH^{p}VH^{q}W)}$ factorize in the same way as in (\ref{eq:factorize_contraction}), since any insertions of the vertex shown in Fig. \ref{fig:Feyn_Diag} that involve factors of $H$ that ``cross" an operator $\mathcal{O}$, will be proportional to $\Tr(\mathcal{O})$.  Therefore,
\begin{align}
\overline{\Tr(H^{n}VH^{m}WH^{p}VH^{q}W)} = {\Tr(\overline{H^{n}}V\overline{H^{m}}W\overline{H^{p}}V\overline{H^{q}}W)}\nonumber
\end{align}
In the large-$N$ limit, the probability distribution for a single eigenvalue of $H$ is known to be \cite{Non_Gaussian}
\begin{align}
\rho(\lambda) = a^{2}\Big(1 + 8g\,a^{2} + 4g\lambda^{2}\Big)\frac{\sqrt{4\,a^{2}-\lambda^{2}}}{2\pi a^{2}}\Theta(2a - |\lambda|)
\end{align}
where the constant $a$ is determined from the coupling constant $g$ by the expression \cite{Non_Gaussian}:
\begin{align}
a^{2} = \frac{-1 + \sqrt{1 + 48g}}{24g}.
\end{align}
We now observe that 
\begin{align}
\overline{\lambda^{2n}} = 2\,a^{2n+2}&\Bigg[\left(4ga^{2} + \frac{1}{2}\right) \frac{(2n)!}{(n!)^{2}(n+1)} \nonumber\\
&+ 2ga^{2}\frac{(2n+2)!}{(n+1)!^{2}(n+2)}\Bigg]
\end{align}
while $\overline{\lambda^{2n+1}} = 0$.  Then, we observe that
\begin{align}
Q(z, g) \equiv \sum_{n=0}^{\infty}\frac{z^{n}\overline{\lambda^{n}}}{n!} = a^{2}(1+8ga^{2})f(az) + 4ga^{4}h(az)\nonumber
\end{align}
where
\begin{align}
&f(z) \equiv \frac{I_{1}(2z)}{z}\\
&h(z) \equiv \frac{2\,I_{2}(2z)}{z^{2}} + \frac{4\,I_{3}(2z)}{z}
\end{align}
and $I_{n}(z)$ is the $n$th modified Bessel function of the first kind.  

From these expressions, and using the fact that $Q(z,g) = Q(-z,g)$, we derive the ensemble-averaged out-of-time-ordered commutator for two operators $V$ and $W$ that are traceless, and disjoint at time $t=0$:
\begin{align}
\overline{\langle V(t)\,W\,V(t)\,W\rangle_{\beta}} = \frac{\Tr(V^{2}W^{2})}{N}\frac{Q(\beta + it,g) Q(it,g)^{3}}{Q(\beta,g)}
\end{align}
And therefore,
\begin{align}
&\overline{C_{\beta}(t)} = \frac{\Tr(V^{2}W^{2})}{N}\left[1 - \mathrm{Re}\left\{\frac{Q(\beta + it,g)}{Q(\beta,g)}\right\}Q(it,g)^{3}\right]\\
&\overline{D_{\beta}(t)} = \frac{\Tr(V^{2}W^{2})}{N}\mathrm{Im}\left\{\frac{Q(\beta + it,g)}{Q(\beta,g)}\right\}Q(it,g)^{3}
\end{align}

We observe that as $g\rightarrow\infty$, $a^{2} = (1/24)\sqrt{48/g} + O(g^{-1})$.  From the form of $Q(z,g)$, we note that $Q(it, g)$ should vanish at a time $t$ when $iat\sim O(1)$.  As a result, the ``scrambling time" -- when the out-of-time-ordered commutator saturates -- \emph{grows} as $t_{s} \sim g^{1/4}$ when $g$ is large.

\section{Behavior of the OTOC for Ensembles with $H\rightarrow UHU^{\dagger}$ Symmetry }
We argue that the OTOC for an ensemble of Hamiltonians invariant under the unitary transformation
\begin{align}
H\,\longrightarrow\,UHU^{\dagger}
\end{align}
where $U$ is any $N\times N$ unitary, has a universal, power-law decay to its asymptotic form.  Our argument proceeds by observing that this ensemble-averaged OTOC may be calculated by first performing the unitary transformation $H\rightarrow UHU^{\dagger}$, and averaging over the choice of unitary matrix $U$ -- which we take to be Haar random, i.e. uniformly distributed over the group of $N\times N$ unitary matrices.  Therefore,  
\begin{align}
&\overline{\langle V(t)W(0)V(t)W(0)}_{\beta} =\nonumber\\
& {\frac{\mathbb{E}_{U}\Tr\left\{Ue^{-(\beta + it)H} U^{\dagger}VUe^{iHt} U^{\dagger}WUe^{-iHt} U^{\dagger}VUe^{iHt} U^{\dagger}W\right\}}{\Tr(e^{-\beta H})}}\nonumber
\end{align}
where $\mathbb{E}_{U}$ denotes the average over the Haar-random unitary $U$.

Performing the unitary average results in an expression of the form
\begin{align}\label{eq:E_U}
&\mathbb{E}_{U}\Tr\left\{Ue^{-(\beta + it)H} U^{\dagger}VUe^{iHt} U^{\dagger}WUe^{-iHt} U^{\dagger}VUe^{iHt} U^{\dagger}W\right\}\nonumber\\
&= \frac{1}{N}\Tr(V^{2}W^{2})\,h_{1}(\beta, t) + \frac{\Tr(VW)^{2}}{N^{2}}\,h_{2}(\beta, t) \nonumber\\
&\,\,\,\,+ \frac{\Tr(V^{2})\Tr(W^{2})}{N^{2}}\,h_{3}(\beta, t)
\end{align}
Here, each function $h_{1}$, $h_{2}$ and $h_{3}$ is a weighted sum of traces of various products of the four operators $e^{\pm iHt}$ and $e^{-(\beta + it)H}$ that appear in (\ref{eq:E_U}); for illustrative purposes, we write out the terms that appear in the expansion of one of the functions 
\begin{align}\label{eq:h1}
h_{1}&(\beta, t) = \frac{c_{1}}{N^{4}}\Tr(e^{iHt})^{2}\Tr(e^{-iHt})\Tr(e^{-(\beta + it)H}) \nonumber\\
&+ \frac{c_{2}}{N^{3}}\Tr(e^{iHt})^{2}\Tr(e^{-(\beta + 2it)H})\nonumber\\
&+ \frac{c_{3}}{N^{3}}\Tr(e^{iHt})\Tr(e^{-iHt})\Tr(e^{-\beta H})\nonumber\\
&+ \frac{c_{4}}{N^{2}}\Tr(e^{iHt})\Tr(e^{-(\beta+ it) H})\nonumber\\
&+ \frac{c_{5}}{N^{3}}\Tr(e^{2iHt})\Tr(e^{-iHt})\Tr(e^{-(\beta + it)H})\nonumber\\
&+ \frac{c_{6}}{N^{2}}\Tr(e^{-iHt})\Tr(e^{-(\beta - it)H}) + \frac{c_{7}}{N}\Tr(e^{-\beta H})
\end{align}
Notably, while the coefficients $\{c_{i}\}$ are all non-zero, they are not all $O(1)$, and a careful treatment of the Haar average is required to extract their $N$-dependence.  For our purposes, a detailed analysis of the $N$-dependence of the coefficients will be unnecessary for extracting the result quoted in the main text.


We assume for simplicity of presentation and without loss of generality, that the density of states of the Hamiltonian, $\rho(\epsilon)$, is an even function on the interval $[-2J, 2J]$.  The function vanishes at the edge of this energy window and outside of this interval.  From this, and from the form of the functions $h_{i}(\beta + it, H)$ given in Eq. (\ref{eq:h1}) we recognize that the time-dependent piece of the OTOC 
\begin{align}
\overline{F_{\beta}(t)} = \overline{\langle V(t)W(0)V(t)W(0)}_{\beta}
\end{align}
may be factorized into the form
\begin{align}
\overline{F_{\beta}(t)}& = \overline{F_{\beta}(\infty)} + \frac{1}{N}{\Tr(e^{iHt})} \,\,\,{{f}(\beta , t)}
\end{align}
where the function ${f(\beta, t)}$ may be written explicitly using Eq. (\ref{eq:E_U}) and (\ref{eq:h1}).  

Notably, since the density of states is an even function over the interval $[-2J, 2J]$, we observe that
\begin{align}
\frac{1}{N}\,&{\Tr(e^{zH})} = -\frac{2}{z} \int_{0}^{2J}d\omega\,\sinh(z\omega)\,\rho'(\omega)\nonumber\\
&\overset{|\epsilon_{0}z|\gg 1}{\sim}\begin{array}{cc}
\displaystyle\frac{\sinh( z \epsilon_{0})}{z\,|z|}\rho'(\epsilon_{0} - |z^{-1}|) 
\end{array}
\end{align}
in the last line, we have used the fact that the dominant contribution to the integral will be from the part of the density of states where the derivative of the density of states $\rho'(\epsilon)$ is most singular.   From this, we conclude that the function ${f(\beta, t)}$ vanishes at long times, at any temperature, and that the OTOC approaches its asymptotic value at least as fast as $N^{-1}\,{\Tr(e^{-iHt})}$, i.e.
 \begin{align}\label{eq:power_law_bound}
\overline{F_{\beta}(\infty)} - \overline{F_{\beta}(t)} \lesssim \frac{\sin( \epsilon_{0}t)}{t^{2}}\rho'(\epsilon_{0} - t^{-1})
\end{align}
From this, we further conclude that the scrambling time is set by the energy scale, at which the derivative of the density of states is most singular $t_{s}\sim \epsilon_{0}^{-1}$.
\end{document}